\documentclass[12pt]{article}

\usepackage{epsf}
\usepackage[dvips]{graphicx}

\begin{document}

\baselineskip 24pt

\begin{center}
{\bfseries DEUTERON-PROTON ELASTIC
SCATTERING AT INTERMEDIATE ENERGIES
}

\vskip 5mm

N.B.Ladygina$^\dag$

\vskip 5mm

{\small
 {\it
Laboratory of High Energies,
Joint Institute for Nuclear Research, 141980 Dubna, Russia,
}
\\
$\dag$ {\it
E-mail: ladygina@sunhe.jinr.ru
}}
\end{center}

\vskip 5mm

\begin{center}
\begin{minipage}{150mm}
\centerline{\bf Abstract}

\baselineskip 24pt

 The deuteron-proton elastic scattering has been studied in the multiple
 scattering expansion formalism. The essential attention has been 
 given to such relativistic problem as a deuteron wave function
 in  a moving frame and transformation of spin states due to Wigner
 rotation.  Parameterization of the nucleon-nucleon $t$-matrix has been used
 to take the off-energy shell effects into account.
 The vector, $A_y,$ and tensor, $A_{yy}$, analyzing powers of 
 the deuteron have been calculated at two deuteron kinetic energies:
 395 MeV and 1200 MeV. The obtained results are compared with 
the experimental data.
\end{minipage}
\end{center}

\vspace{1cm}{\bf PACS:} 21.45.+v, 25.45.-z, 25.45.De, 24.10.Jv, 24.70.+s

\vspace{1cm}{\it Keywords:} Few-body system, Deuteron-induced reactions,
Elastic scattering,
Polarization in nuclear reactions

\section{Introduction}

The study of the deuteron-proton elastic scattering has a longtime
story both from the experimental \cite{395}- \cite{ghaz} and theoretical 
\cite {glrew}- \cite {bles} (and refs. therein) points
of view. This process is the simplest 
example of the hadron nucleus collision that is why the interest to 
this reaction is justified. A number  of experiments 
on deuteron- nucleon elastic scattering is aimed at getting some information
about the deuteron wave function as well as nucleon-nucleon amplitudes
from $Nd$ scattering observables. Moreover, the study of the reaction 
mechanisms,
investigations of the few-body scattering dynamics are also very 
important.

A good theoretical description of the deuteron-nucleon process is 
obtained for low energies, where the multiple scattering 
formalism 
based on the solution of the Faddeev equations has been applied to this
problem   \cite {glrew}. Nevertheless, this method requires the realistic 
nucleon-nucleon
potential but the well known $NN$  potentials such as Bonn \cite{B}, CD-Bonn
\cite {cd},
 Paris \cite {NN}, Nijmigen \cite {nijm}, AV18 \cite {av18}, can be used
only  up to about 300-350 MeV $NN$ laboratory kinetic
energy. This fact  indicates a necessity to develop new relativistic
 nucleon-nucleon potentials which describe existing $NN$ data up to
 1 GeV and higher. Many efforts to get these kind of $NN$ 
 potentials have been undertaken
 \cite {el1}-\cite {khokhlov} but these potentials describe very well only
 low $NN$ phase shifts. It is possible in the future more acceptable
relativistic $NN$ potential will be developed what allows to extend
energy region where Faddeev equations formalism can be applied.

At present there are  experimental data on the $dp$
elastic scattering in a wide range, where the deuteron kinetic
energy is above 400 MeV  \cite{395}- \cite{ghaz} (and refs. therein).
Already from a few hundred MeV of nucleon-nucleon interaction energy
the relativistic effects begin to play a very  important role. A relativistic
quantum theory including relativistic kinematics, Lorentz transformation
and the relativistic spin theory should be applied to consider a few-body
scattering. The basic principles of this theory were formulated in Ref.
 \cite {sucher} 
where results of the works of Dirac \cite {dirac}, Wigner \cite {wigner},
Foldy \cite {foldy}, etc., were developed. In Ref. \cite {polyz} review devoted
to the relativistic dynamics is presented.
Recently relativistic kinematics, boost effects and Wigner rotation
have been included into the three-nucleon Faddeev equations \cite {el4}, \cite {el5}.
 
The present paper has realized the semi-phenomenological
method based on the parameterizations of the deuteron wave function (DWF)
and nucleon-nucleon vertices. At intermediate 
energies the relative momentum of the two nucleons in the deuteron is 
rather large and, in general, the use of a relativistic deuteron wave function
 is necessary.
In Refs. \cite {gross}, \cite {tjon} these DWF's were obtained by
solution of the Blankenbecler-Sugar equation. But in this approach
one of the nucleons is on-energy shell that limits the employment of 
these DWF's. The two nucleons in the deuteron are considered as the off-energy
shell particles in Ref. \cite {kaptari}, where the deuteron
wave function is the result of the Bethe-Salpeter equation solution.
The importance of  the relativistic DWF application to describe the
deuteron-proton backward elastic scattering was  demonstrated in this paper.  

However, at the energies considered in the present paper it is 
possible to use nonrelativistic DWF such as Paris \cite {par}, Bonn \cite {B},
CD-Bonn \cite {cd} due to transition into the deuteron Breit frame.
It allows one to decrease the two nucleon relative momentum both in the initial
and final deuterons. In this paper the CD-Bonn DWF has been used as a
more modern one. It should be noted that the Lorentz transformation and
relativistic spin theory have been applied to describe the transformation
from the deuteron rest frame to the Breit system. As a result, the deuteron
wave function has a more complicated spin structure, than usual nonrelativistic
DWF and depends on the two variables. The analogous result was obtained in Refs. 
\cite {karmanov1}-\cite {carbonell} for the DWF on the light front.

The model offered in Ref.\cite {alberi} to investigate
the $dp$-elastic scattering at intermediate energies, has been developed 
further in the present paper. First of all, the deuteron wave function 
transformation has been done. Secondly,  
the phase shift analysis data are not used to describe
the nucleon-nucleon interaction in comparison with Ref. \cite {alberi},
where
the energy dependence in $NN$-vertices
 important for correct integrations, is lost. Besides, the phase shift
analysis gives information about on-energy shell nucleons.
But in our method the off-energy shell effects have to be taken into 
consideration and it 
 can be done by  using  some parameterization
describing  the nucleon-nucleon interaction. Thus, 
the $NN$ $t$ -matrix constructed by Love and Franey
\cite {LF} has been used in our calculations.

The presented method has been applied to calculate two simplest
polarization observables: vector, $A_y$, and tensor,
$A_{yy}$, analyzing powers at two deuteron kinetic energies:
395 MeV and 1200 MeV.
The results of these calculations are compared with the
experimental data obtained in Saclay \cite {395}, \cite {1200} and 
Argonne \cite{1200a}. It should be noted that these observables and 
energies are examples of the presented method employment. In the future 
this method is planned  to describe other polarization observables at 
deuteron kinetic  
energies between 400 and 1200 MeV.

The paper is organized as follows. Section 2 gives the general
formalism: the transformation of the
deuteron wave function from the rest frame to a 
moving deuteron system, is considered; the
description of the nucleon-nucleon interaction is presented;
the calculations of the scattering amplitude terms
are performed. Section 3 defines
 the polarization observables, $A_y$ and $A_{yy}$.
 The results of the calculations are
discussed in Sect.4. The conclusions are contained in Sect.5.

\section{General formalism}

 According to the three-body collision theory,
the amplitude of the deuteron-proton elastic scattering $\cal J$
is defined by the matrix element of the transition operator $U_{11}$:
\begin{eqnarray}
U_{dp\to dp}=\delta(E_d+E_p-E^\prime_d-E^\prime _p) {\cal J}=
<1(23)|[1-P_{12}-P_{13}]U_{11}|1(23)>~~.
\end{eqnarray}
Here the state $|1(23)>$ corresponds to the configuration, 
when  nucleons 2 and 3
form the deuteron state and nucleon 1 is free. 
Appearance of the permutation operators for two nucleons $P_{ij}$
reflects the fact that the initial and final states are antisymmetric
due to exchange of the two particles.

The transition operators for rearrangement scattering 
 are defined by the Alt-Grassberger-Sandhas equations
 \cite{Alt}, \cite{AGS}
\begin{eqnarray}
U_{11}&=&~~~~~~~~t_2g_0U_{21}+t_3g_0U_{31}
\nonumber\\
U_{21}&=&g_0^{-1}+t_1g_0U_{11}+t_3g_0U_{31}
\\
U_{31}&=&g_0^{-1}+t_1g_0U_{11}+t_2g_0U_{21}~,
\nonumber
\end{eqnarray}
where $t_1=t(2,3)$,etc., is the t-matrix of the two-nucleon interaction and
$g_0$ is the free three-particle propagator. The indices $ij$ for the
transition operators $U_{ij}$ denote free particles $i$ and $j$ in the final
 and initial states, respectively.

Iterating these equations up to the $t_i$ second order terms  we can present
the reaction amplitude as a sum of the following three contributions:
 one nucleon exchange,
single scattering and double scattering terms, --
\begin{eqnarray}
\label{contrib}
{\cal J}_{dp\to dp}&=&{\cal J}_{ONE}+{\cal J}_{SS}+{\cal J}_{DS}
\nonumber\\
{\cal J}_{ONE}&=&-2<1(23)|P_{12}g_0^{-1}|1(23)>
\nonumber\\
{\cal J}_{SS}&=&2<1(23)|t_3^{sym}|1(23)>
\\
{\cal J}_{DS}&=&2<1(23)|t_3^{sym}g_0t_2^{sym}|1(23)>~~,
\nonumber
\end{eqnarray}
where we have introduced notations for antisymmetrized operators
$t_2^{sym}=[1-P_{13}]t_2$  and $t_3^{sym}=[1-P_{12}]t_3$. 
Before we consider each of these elements in detail, we must
determine the reference frame.

\vspace{1cm}
{\bf {\large Breit system}}
\vspace{0.5cm}

 The most popular deuteron wave functions are well known only in 
 the nonrelativistic region.
In this connection it is very important to
choose the frame, where the application of the nonrelativistic deuteron
wave function is possible.
The frame which minimizes the relative momenta of the nucleons in the both
deuteron wave functions, is the deuteron Breit system. In this frame
the deuterons move  in opposite directions with equal momenta (fig.1).
As a consequence, the energies of the initial and final deuterons (and protons)
are equal to each other.
\begin{eqnarray}
&&E_d=E^\prime_d=\sqrt{M_d^2+\vec Q^2}, ~~~~~~~~~
E_p=E^\prime_p=\sqrt{m^2+\vec p^2}
\nonumber\\
&&(\vec p\vec Q)=-\vec Q^2~~.
\end{eqnarray}

We take the orthonormal basis
\begin{eqnarray}
\vec z=\frac {\vec p-\vec p^\prime}{|\vec p-\vec p^\prime|}=-\hat Q,~~~~~
\vec x=\frac {\vec p+\vec p^\prime}{|\vec p+\vec p^\prime|}=
\widehat{p+Q},~~~~~\vec y=\vec z\times \vec x=\hat p\times \hat Q ~~,
\end{eqnarray}
 where $z$-axis goes along 
the transfer momentum and
$x$-axis $-$ along the average momentum of the initial and final
protons. As usual, the $y$-axis is normal to
the scattering plane.

\vspace{1cm}
{\bf\large Deuteron wave function}
\vspace{0.5cm}

In order to get the wave function of the moving deuteron,
let us consider an arbitrary frame where two nucleons have momenta $\vec p_1$,
 $\vec p_2$ and energies $E_1=\sqrt{m_N^2+\vec p_1\- ^2}$, 
$E_2=\sqrt{m_N^2+\vec p_2\- ^2}$, respectively. Here $m_N$ is the nucleon mass.
 These momenta are related
with those in the center-of-mass (c.m.) by Lorenz transformation
\begin{eqnarray}
&&L(\vec u) p_1= (E^*,\vec p)
\nonumber\\
&&L(\vec u) p_2=(E^*, -\vec p)
\end{eqnarray}
with velocity 
\begin{eqnarray}
\vec u=\frac{\vec p_1+\vec p_2}{E_1+E_2}~~.
\end{eqnarray}
Here  the c.m. energy of one of the  nucleons $E^*$ is related with
Mandelstam variable $s$ by
\begin{eqnarray}
E^*=\sqrt s/2~~.
\end{eqnarray}
Thus, we can introduce  new variables $\vec Q$ and $\vec k$ 
which can be expressed through $\vec p_1$
and $\vec p_2$ 
\begin{eqnarray}
\label{mom}
&&\vec Q=\vec p_1 +\vec p_2
\nonumber\\
&&\vec k=\frac{(E_2+E^*)\vec p_1-(E_1+E^*)\vec p_2}{E_1+E_2+2E^*}~~.
\end{eqnarray}
Then a two-nucleon state in the $(\vec p_1, \vec p_2)$ system is connected
with a two-nucleon  state in the center-of-mass by the relation
\begin{eqnarray}
\label{p1p2}
|\vec p_1,\vec p_2>=J^{-1/2}(\vec p_1,\vec p_2)
W_{1/2} (\vec p_1,\vec u)W_{1/2} (\vec p_2,\vec u)
|\vec k,\vec Q>~~,
\end{eqnarray}
where $\vec k$ is the relative momentum  of the two nucleons in the c.m.
The normalization factor provides conditions of orthonormality and completeness
  and 
 is defined by the Jacobian of the transformations
\cite {sucher}
\begin{eqnarray}
\label{Jac}
J(\vec p_1,\vec p_2)=\frac{\partial (\vec p_1, \vec p_2)}
{\partial (\vec k,\vec Q)}=\frac{2}{E^*}\frac{E_1E_2}{E_1+E_2}~~.
\end{eqnarray}
The Wigner rotation operator in the spin space of the $i$-th nucleon
has the standard form
\begin{eqnarray}
\label{w12}
W_{1/2}(\vec p_i, \vec u)=exp\left\{ -i\omega_i(\vec n_i\vec\sigma_i)/2\right\} =
cos(\omega_i/2)[1-i(\vec n_i\vec\sigma_i)tg(\omega_i/2)]
\end{eqnarray}
with the rotation axis
\begin{eqnarray}
\label{axis}
\vec n_i=\frac{\vec u\times \vec p_i}{|\vec u\times \vec p_i|}
\end{eqnarray}

Ref.\cite {sucher} shows that  under a pure Lorenz transformation 
\begin{eqnarray}
&&L(\vec u^\prime) \vec P=\vec Q
\end{eqnarray}
the wave function of the 
bound state transforms
in the same way as the state of a single particle
\begin{eqnarray}
W(L_{\vec u^\prime})|\vec P>=
\sqrt {\frac{E_{\vec Q}}{E_{\vec P}}}~~W_1(\vec Q, \vec u^\prime )|\vec Q>~~,
\end{eqnarray}
where the $W_1$ is the Wigner rotation operator for spin $1$ particle 
\begin{eqnarray}
W_1(\vec Q, \vec u^\prime)=exp\left\{ -i\omega(\vec n^\prime\vec S)\right\}
\end{eqnarray}
with the rotation axis
\begin{eqnarray}
\vec n^\prime =\frac{\vec u^\prime\times \vec Q}{|\vec u^\prime\times \vec Q|}~~.
\end{eqnarray}

The deuteron wave function in the rest frame of the deuteron 
depends on only one variable $\vec k$, which is the
 relative momentum of the  outgoing proton and neutron
\begin{eqnarray}
\label{dwf0}
<m_p m_n|\Omega_d|{\cal M}_d>=
\frac{1}{\sqrt{4\pi}}<m_p m_n|\left\{ u(k)+\frac{w(k)}{\sqrt 8}
[3(\vec\sigma_1 \hat k)(\vec\sigma_2\hat k)-(\vec \sigma_1\vec\sigma_2)]
\right\}|{\cal M}_d>~~,
\end{eqnarray}
where $u(k)$ and $w(k)$ describe the $S$ and $D$ components of 
the deuteron wave function   \cite{B}, \cite{cd}, \cite{par}, $\hat k$ is 
the unit vector in $\vec k$ direction and ${\cal M}_d, m_p, m_n$ are 
spinors of the deuteron, proton, and neutron, respectively.

Obviously, the deuteron rest frame corresponds to the center-of-mass
of the two outgoing nucleons, $\vec P=\vec p_n+\vec p_p=\vec 0$.
 Then the proton -neutron relative momentum 
$\vec k$  in Eq.(\ref {dwf0})
 is related with the nucleon momenta $\vec p_1$ and $\vec p_2$
in the arbitrary frame by Eq.(\ref {mom}).
Using the formulae (\ref {mom})-(\ref {dwf0}) we get the expression for
the deuteron wave function in the frame, where the deuteron is
moving and its momentum is  $\vec Q$
\begin{eqnarray}
\label{reldwf}
\hspace{-0.5cm}
<\vec p_1~ \vec p_2, m_1m_2|\Omega_d |\vec Q,{\cal M}_d>&=&
\sqrt{\frac{M_d}{E_{\vec Q}}}~
\sqrt{\frac {E^*(E_1+E_2)}{2E_1E_2}}
cos\frac{\omega_1}{2}cos\frac{\omega_2}{2}\cdot
\\
&&\hspace{-1cm}
<\vec k\vec Q, m_1^\prime m_2^\prime|
[1-i(\vec n\vec\sigma_1)tg\frac{\omega_1}{2}]
[1+i(\vec n\vec\sigma_2)tg\frac{\omega_2}{2}]
\Omega_d
|\vec 0, {\cal M}_d>~~.
\nonumber
\end{eqnarray}
The Wigner rotation is performed  around the normal to the
$(\vec k\vec Q)$  plane 
\begin{eqnarray}
\label{daxis}
\vec n=\frac{\vec p_1\times\vec p_2}{|\vec p_1\times\vec p_2|}=
\frac{\vec k\times\vec Q}{|\vec k\times\vec Q|}
\end{eqnarray}
on the rotation angles, which are defined by the relations \cite {ritus}:
\begin{eqnarray}
\label{angles}
&&tg\frac{\omega_1}{2}=\frac{|\vec p_1\times\vec p_2|}
{m_N(E_1+E_2+2E^*)+2E^*(E_1+E^*)}
\\
&&tg\frac{\omega_2}{2}=\frac{|\vec p_1\times\vec p_2|}
{m_N(E_1+E_2+2E^*)+2E^*(E_2+E^*)}~~.
\nonumber
\end{eqnarray}

The expression (\ref {reldwf}) can be presented through six terms:
\begin{eqnarray}
<\vec p_1~ \vec p_2, m_1m_2|\Omega_d |\vec Q,{\cal M}_d>&=&
g_1(\vec k,\vec Q)+
g_2(\vec k,\vec Q)(\vec \sigma_1\vec n)(\vec \sigma_2\vec n)+
g_3(\vec k,\vec Q)(\vec \sigma_1\vec \sigma_2)+
\nonumber\\
&+&g_4(\vec k,\vec Q)(\vec \sigma_1\hat k)(\vec \sigma_2\hat k)+
g_5(\vec k,\vec Q)[(\vec \sigma_1+\vec \sigma_2)\vec n]+
\\
&+&g_6(\vec k,\vec Q)[(\vec \sigma_1\hat k)(\vec \sigma_2\vec n\times\hat k)+
(\vec \sigma_1\vec n\times\hat k)(\vec \sigma_2\hat k)].
\nonumber
\end{eqnarray}

Solving a relativistic equation one can obtain functions $g_i$ as it was
done in Ref.\cite {karmanov1}-\cite {carbonell} on the light front.
But in this paper a usual nonrelativistic DWF is taken as input.
Therefore, functions $g_i$ are defined as  linear combinations of $u$ and
$w$ ($S$- and $D$-waves).

This way we have got the relation between the deuteron wave function
in two systems: the deuteron rest frame and  the arbitrary moving system. 
Note, the Wigner rotation corrections are small at the energies under 
consideration, but momenta transformations are very important since they
give us an opportunity to use the nonrelativistic deuteron wave functions
in a rather wide energy region.

\vspace{1cm}
{\bf\large One nucleon exchange scattering}
\vspace{0.5cm}

The first contribution into the $dp$-elastic scattering amplitude ${\cal J}$
in Eq.(\ref {contrib}) is the  one nucleon exchange (ONE) term.
The corresponding diagram is presented in Fig.1a. Applying the
definitions of the wave function of a moving deuteron and three-
nucleon free propagator, we can write ONE amplitude in the following 
form:
\begin{eqnarray}
\label{one}
{\cal J}_{ONE}=&-2&_{1(23)}<\vec p^\prime m^\prime \tau^\prime; 
-\vec Q {\cal M}_d^\prime 0|\Omega_d^\dagger (23) P_{12}
\\
&&(E_d+E_p-\hat K_1-\hat K_2-\hat K_3 +i\varepsilon )
\Omega_d(23)|
\vec Q {\cal M}_d 0;\vec p m \tau>_{1(23)}~~,
\nonumber
\end{eqnarray}
where $m$, $m^\prime$ are the spin projections of the initial and final protons,
$\tau$, $\tau^\prime$ are their isospin projections, respectively.  
The kinetic-energy operator has the standard definition,
 $\hat K_i|\vec p_i>=\sqrt{m_N^2+\vec p_i\-^2}|\vec p_i>$ .

After a straightforward calculation we have the expression
for the ONE amplitude
\begin{eqnarray}
{\cal J}_{ONE}&=&-\frac{1}{2}(E_d-E_p-\sqrt{m_N^2+\vec p~^2-\vec Q\-^2})
\cdot
\\
&&<\vec p^\prime m^\prime;-\vec Q {\cal M}_d^\prime|
\Omega^\dagger_d(23)
[1+(\vec\sigma_1\vec\sigma_2)]
\Omega_d(23)|
\vec Q {\cal M}_d;\vec p m >~~,
\nonumber
\end{eqnarray}
where the definition of  the permutation
operator in spin space 
 $P_{12}(\sigma )=\frac{1}{2}[1+(\vec\sigma_1\vec\sigma_2)]$ has been applied.
The two nucleons relative momentum in the deuteron rest frame
is the argument of the deuteron wave function in Eq.(\ref {dwf0}).
Using Eq.(\ref {mom}) we can express these variables for the initial
 $\vec p_0$ and final $\vec p_0^\prime $ deuterons, correspondingly, 
\begin{eqnarray}
\vec p_0&=&\vec p +\vec Q\left[ 1+\frac{E_n+E^*}{E_p+E_n+E^*}\right]
\\
\vec p_0^\prime &=&\vec p +\vec Q\left[ 1-\frac{E_n+E^*}{E_p+E_n+E^*}\right]~~.
\nonumber
\end{eqnarray}
Here $E_n=\sqrt{m_N^2+\vec p~^2-\vec Q\-^2}$ 
 and $E^*=\sqrt{(E_p+E_n)^2-\vec Q\-^2}/2$ 
are the struck neutron energy in the moving deuteron frame and rest deuteron
frame, respectively. Note, that $|\vec p_0|=|\vec p_0^\prime |$.
The Wigner rotations are performed around $y$-axis
on the equal angles 
both for the initial and final deuterons. Using Eq.(\ref {angles}) we get
\begin{eqnarray}
tg\frac{\omega_1}{2}=tg\frac{\omega_1^\prime}{2}=
\frac{|\vec p\times\vec Q|}{m_N(E_p+E_n+2E^*)+2E^*(E_p+E^*)}
\\
tg\frac{\omega_2}{2}=tg\frac{\omega_2^\prime}{2}=
\frac{|\vec p\times\vec Q|}{m_N(E_p+E_n+2E^*)+2E^*(E_n+E^*)}~~,
\nonumber
\end{eqnarray}
where the angles $\omega_1$, $\omega_1^\prime$ are related with
the projectile and outgoing protons, respectively, while 
$\omega_2$, $\omega_2^\prime$ correspond to the struck neutron
(Fig.1a).

\vspace{1cm}
{\bf\large Single scattering}
\vspace{0.5cm}

The other term in the $dp$-elastic scattering amplitude 
Eq.(\ref {contrib}) is the single scattering one. 
The corresponding diagram is presented in Fig.(1b).
Following the  standard procedure and applying the definition of the
moving deuteron wave function Eq.(\ref {reldwf}), we get the expression 
for the single scattering amplitude:
\begin{eqnarray}
\label{ss}
{\cal J}_{SS}&=&\int d\vec q~^\prime <-\vec Q {\cal M}_d^\prime |
\Omega_d^\dagger|
\vec q~^\prime m^{\prime\prime}, -\vec Q-\vec q~^\prime m_3^\prime>
\\
&&\hspace{-1cm}
<\vec p~^\prime m^\prime, -\vec Q-\vec q~^\prime |
\frac{3}{2}t^1_{12}+\frac{1}{2}t^0_{12}|
\vec p m, \vec Q -\vec q~^\prime m_2^\prime >
<\vec q~^\prime m^{\prime\prime},\vec Q-\vec q~^\prime m_2^\prime|
\Omega_d|\vec Q {\cal M}_d>~~.
\nonumber
\end{eqnarray}
The relative momenta of two nucleons for the initial and final 
deuterons are
\begin{eqnarray}
\vec p_0&=&\vec q~^\prime -\vec Q\frac{E_2+E^*}
{E_2+E_3+2E^*}
\\
\vec p_0^\prime &=&\vec q~^\prime +\vec Q
\frac{E_2+E^{\prime *}}
{E_2+E_3^\prime +2E^{\prime *}}~~,
\nonumber
\end{eqnarray}
where the nucleons energies $E_2$, $E_3$, $E_3^\prime $ 
in the reference frame are defined by the 
standard manner (Fig.1b)
\begin{eqnarray}
\label{e2e3}
E_2=\sqrt{m_N^2+\vec q~^{\prime 2}}~,
~~~~~~
E_3=\sqrt{m_N^2+(\vec Q -\vec q~^\prime)^2}~,
~~~~~~
E_3^\prime =\sqrt{m_N^2+(\vec Q +\vec q~^\prime)^2}
\end{eqnarray}
and these energies in the center-of-mass of the two nucleons
forming the initial and final deuterons
 are equal, correspondingly, to
\begin{eqnarray}
E^*=\frac{1}{2}\sqrt{(E_2+E_3)^2-\vec Q^2}~,
~~~~~~~
E^{\prime *}=\frac{1}{2}\sqrt {(E_2+E_3^\prime )^2-\vec Q^2}~~.
\end{eqnarray}

The Wigner rotation axis goes along the normal to $(\vec q~^\prime \vec Q)$ plane
\begin{eqnarray}
\vec n=\frac{\vec q~^\prime\times \vec Q }{|\vec q~^\prime\times \vec Q |}
\end{eqnarray}
and Wigner rotation angles are defined in accordance with 
Eq.(\ref {angles}) for the initial
\begin{eqnarray}
\label{tan}
tg\frac{\omega_2}{2}&=&\frac{|\vec q~^\prime\times \vec Q |}
{m_N(E_2+E_3+2E^*)+2E^*(E_2+E^*)}
\\
tg\frac{\omega_3}{2}&=&-\frac{|\vec q~^\prime\times \vec Q |}
{m_N(E_2+E_3+2E^*)+2E^*(E_3+E^*)}
\nonumber
\end{eqnarray}
and final deuterons
\begin{eqnarray}
tg\frac{\omega_2^\prime}{2}&=&-\frac{|\vec q~^\prime\times \vec Q |}
{m_N(E_2+E_3^\prime+2E^{\prime *})+2E^{\prime *}(E_2+E^{\prime *})}
\\
tg\frac{\omega_3^\prime}{2}&=&\frac{|\vec q~^\prime\times \vec Q |}
{m_N(E_2+E_3^\prime+2E^{\prime *})+2E^{\prime *}(E_3^\prime +E^{\prime *})}~~.
\nonumber
\end{eqnarray}

Further we have to give essential attention to the two nucleon
$t^i$-matrix in Eq.({\ref{ss}}). Here index $i$ denotes the isospin state
of the nucleon-nucleon system. This $t$-matrix 
is the half-off-energy shell one.
 Since we assume to use the formalism presented here to
 investigate  the processes at rather high energies, 
we can not 
define the high-energy $NN$ $t$-matrix solving the 
Lippmann-Schwinger equation with any known potential.
In this situation it is possible to apply some parameterization
as  in Refs.\cite {LSh}, \cite {epj} to study
 the $dp$ breakup in the GeV region.

\vspace{1cm}
{\bf\large Nucleon-nucleon $t$-matrix}
\vspace{0.5cm}

In order to describe the nucleon- nucleon  interaction in a wide energy
region, we use the Love and Franey parameterization \cite {LF}.
In this model the on-shell $NN$ $t$-matrix
 is defined in the center-of-mass
\begin{eqnarray}
\label{tnn}
<\kappa^\prime  m_1^\prime m_2^\prime |t|
\kappa m_1m_2>
&=&<\vec\kappa^\prime  m_1^\prime m_2^\prime |
A+B(\vec\sigma_1 \hat N^*)(\vec\sigma_2 \hat N^*)+
C(\vec\sigma_1 +\vec\sigma_2 )\cdot \hat N^* +
\nonumber\\
&&\hspace{2cm}
D(\vec\sigma_1 \hat q^*)(\vec\sigma_2 \hat q^*) +
F(\vec\sigma_1 \hat Q^*)(\vec\sigma_2 \hat Q^*)
|\vec\kappa m_1m_2>~~,
\nonumber
\end{eqnarray}
where the orthonormal basis is  combinations of the nucleons
relative momenta in the initial $\vec\kappa $ and final $\vec\kappa^\prime$
states
\begin{equation}
\hat q^*=\frac {\vec \kappa -\vec \kappa^\prime}
{|\vec \kappa -\vec \kappa^\prime|}~~,~~
\hat Q^*=\frac {\vec \kappa +\vec \kappa^\prime }
{|\vec \kappa +\vec \kappa^\prime|}~~,~~
\hat N^*=\frac {\vec \kappa  \times \vec \kappa^\prime }{|\vec \kappa
\times\vec \kappa^\prime |}~~.
\end{equation}

We can not directly use this definition of the $t$-matrix
in our calculation. First of all, because this $t$-matrix is defined 
in the c.m.,
while we need that in the deuteron Breit frame. Second, we need
the off-energy shell $t$-matrix, but the Love and Franey parameterization
gives us the on-energy shell  matrix. However, the both  problems can be
solved.

Using formulae (\ref {p1p2})-(\ref {axis}) we can connect expressions for
the $NN$ $t$-matrix in the c.m. with the same matrix in the reference 
frame by the following relation:
\begin{eqnarray}
\label{tlab}
<\vec p^\prime \vec p_3^\prime; m^\prime m_3^\prime|t|
\vec p \vec p_3; m m_3>=
{\cal N} <\kappa^\prime  m_1^\prime m_2^\prime |W^\dagger_{1/2 } (\vec p^\prime )
W^\dagger_{1/2 }(\vec p_3^\prime )
~t~ W_{1/2} (\vec p)
W_{1/2 }(\vec p_3)|
\kappa m_1m_2>~~.
\end{eqnarray}
Here momentum notations correspond to the ones given in Fig.1b.
The Wigner rotations of the initial and final states are performed
around $\vec \mu$ and $\vec \mu^\prime$ -axes, respectively:
\begin{eqnarray}
\vec \mu=\frac{\vec p\times \vec p_3}{|\vec p\times \vec p_3|},~~~~~~~~
\vec \mu^\prime=\frac{\vec p~^\prime\times \vec p_3~^\prime }
{|\vec p~^\prime \times \vec p_3~^\prime |}~~.
\end{eqnarray}
We introduce the following notations for the Wigner angles:
$\chi_1$, $\chi_1^\prime$ correspond to the projectile and
outgoing protons, while $\chi_3$, $\chi_3^\prime$
are related with the nucleons from the initial and final deuterons,
correspondingly. These angles can be expressed through their tangents 
using Eq.(\ref {angles}):
\begin{eqnarray}
tg\frac{\chi_1}{2}&=&\frac{|\vec p\times \vec p_3|}
{m_N(E_p+E_3+2E^*_{13})+2E^*_{13}(E_p+E^*_{13})}
\nonumber\\
tg\frac{\chi_3}{2}&=&-\frac{|\vec p\times \vec p_3|}
{m_N(E_p+E_3+2E^*_{13})+2E^*_{13}(E_3+E^*_{13})}
\\
tg\frac{\chi_1\-^\prime}{2}&=&\frac{|\vec p~^\prime\times \vec p_3\-^\prime|}
{m_N(E_p+E_3^\prime+2E^{*\prime }_{13})+
2E^{*\prime }_{13}(E_p+E^{*\prime}_{13})}
\nonumber\\
tg\frac{\chi_3\-^\prime }{2}&=&-\frac{|\vec p~^\prime \times \vec p_3\-^\prime |}
{m_N(E_p+E_3^\prime+2E^{*\prime }_{13})+
2E^{*\prime }_{13}(E_3^\prime+E^{*\prime }_{13})}~~.
\nonumber
\end{eqnarray}

As it is shown in ref.\cite {alberi} the spin rotation  corrections to the 
$NN$ $t$- matrix are  very small at the nucleon kinetic energy of 800 MeV .
Therefore in our calculations we neglect the second term in Eq.(\ref {w12}).
Thus, only  cosines of the Wigner angles
are  the Wigner operators in Eq.(\ref {tlab}) and, therefore, the
$NN$ $t$- matrix in the reference frame
 has the same spin structure as that in the c.m. 

The normalization factor in Eq.(\ref {tlab}) is related with the Jacobian of
the transformations Eq.(\ref {Jac}) and equals to
\begin{eqnarray}
{\cal N}=\sqrt{\frac{E_{13}^*(E_p+E_3)}{2E_p E_3}}
\sqrt{\frac{E_{13}^{*\prime }(E_p+E_3^\prime)}{2E_p E_3^\prime}}~~,
\end{eqnarray}
where $E_{13}^*$ and $E_{13}^{\prime *}$ are the initial and final c.m. energies  
of the two nucleons taking part in the interaction, respectively:
\begin{eqnarray}
E^*_{13}=\frac{1}{2}\sqrt{(E_p+E_3)^2-\vec K^2}~,~~~~~~~
E^{*\prime }_{13}=\frac{1}{2}\sqrt{(E_p+E_3^\prime )^2-\vec K^2}~~,
\end{eqnarray}
with a full momentum
\begin{eqnarray}
\vec K=\vec p+\vec Q-\vec q^\prime~~.
\nonumber
\end{eqnarray}

In our calculations we need the half-off-energy shell and off-energy shell
nucleon-nucleon $t$-matrices, while Eq.(\ref{tnn}) gives us only
the on-energy shell matrix.
In order to reach this purpose, we use the result of the 
Ref. \cite {Garc77} which shows the off-energy shell
$t$-matrix to be connected with the on-energy shell matrix by the relation
\begin{eqnarray}
\label{off}
t(E_{onshell};\vec K; \vec \kappa^\prime \vec\kappa)=
F t(\sqrt{s};\vec\kappa^\prime \vec\kappa)~~,
\end{eqnarray}
where the kinematical factor $F$ is defined as
\begin{eqnarray}
\label{F}
F=\frac{E_{onshell}}{\sqrt{s}}\frac{\sqrt{s}+2E^*}{E_{onshell}+E_p+E_3}
\frac{\sqrt{s}+2E^{\prime *}}{E_{onshell}+E_p+E_3^\prime }~~.
\end{eqnarray}
Here  $E_{onshell}$ denotes the energy, which is determined by the following 
condition:
\begin{eqnarray}
\label{eon}
E_{onshell}=E_p+E_3=E_p+E_3^\prime ~~.
\end{eqnarray}
For the single scattering this condition  will be satisfied
at $(\vec Q\vec q^\prime)=0$. Then, 
$E_{onshell}=E_p+\sqrt{m_N^2+\vec Q^2+\vec q~^{\prime 2}}$.

The argument of the on-energy shell $t$-matrix in Eq.(\ref {off}) is
the square root of the 
Mandelstam variable $s$:
\begin{eqnarray}
s=E_{onshell}^2-\vec K^2~~.
\end{eqnarray}

Thus, using the scenario presented here  we can  extend
the parameterized on-energy shell $NN$ $t$-matrix to 
 the  off- energy shell case.  

\vspace{1cm}
{\bf\large Double scattering}
\vspace{0.5cm}

The diagram of the double scattering is shown in Fig.(1c).
The corresponding amplitude ${\cal J}_{DS}$ can be written as
\begin{eqnarray}
\label{ds}
{\cal J}_{DS}&=&N N^\prime\int d\vec q~ ^\prime\int d\vec q^{\prime\prime }
<-\vec Q {\cal M}_d^\prime|\Omega_d^\dagger |
\vec q~^\prime m_2^\prime,\vec Q -\vec q~^\prime~ m_3^\prime>
\\
&&
<m^\prime m_2^\prime m_3^\prime|\Bigg\{ 
t_{12}^1(\sqrt{s_{12}^\prime},\vec\kappa, \vec\kappa ~^\prime ) 
t_{13}^1(\sqrt{s_{13}},\vec k, \vec k ^\prime )+
\nonumber\\
&&
\frac{1}{4}[t_{12}^1(\sqrt{s_{12}^\prime},\vec\kappa, \vec\kappa ~^\prime)+
t_{12}^0(\sqrt{s_{12}^\prime},\vec\kappa, \vec\kappa ~^\prime)]
[t_{13}^1(\sqrt{s_{13}},\vec k, \vec k^\prime )+
t_{13}^0(\sqrt{s_{13}},\vec k, \vec k^\prime )]\Bigg\}
\nonumber\\
&&
\frac{1}{E_d+E_p-E_1^\prime -E_2^\prime -E_3^\prime +i\varepsilon}|m m_2 m_3>
<-\vec Q-\vec q~^{\prime\prime}~m_2,\vec q~^{\prime\prime}~ m_3|\Omega_d|{\cal M}_d>~~.
\nonumber
\end{eqnarray}
As before, we define the relative momenta of the nucleons which
form the initial and final deuterons, respectively,
 in accordance with Eq.(\ref {mom})
\begin{eqnarray}
\vec p_0&=&\vec q~^\prime -\vec Q\frac{E_2+E^*}
{E_2+E_3+2E^*}
\\
\vec p_0\- ^\prime &=&-\vec q~^{\prime\prime} -\vec Q
\frac{E_3^\prime+E^{\prime *}}
{E_2^\prime +E_3^\prime +2E^{\prime *}}~~,
\nonumber
\end{eqnarray}
where the center-of-mass energies of these nucleons
  are related with those in the reference frame by
\begin{eqnarray}
E^*=\frac{1}{2}\sqrt{(E_2+E_3)^2-\vec Q^2},~~~~~~~
E^{\prime *}=\frac{1}{2}\sqrt{(E_2^\prime +E_3^\prime )^2-\vec Q^2}~~.
\end{eqnarray}
Here the energies $E_2$, $E_3$ correspond to the nucleons in the initial
deuteron. The expressions for them are identical to the ones in the single
scattering case and 
 defined by Eq.(\ref {e2e3}). The primed energies
$E_1^\prime, E_2^\prime, E_3^\prime$  are related with the 
corresponding momenta (Fig.1c) in the  standard manner
\begin{eqnarray}
E_1^\prime =\sqrt{m^2+(\vec p+\vec Q -\vec q~^\prime -\vec q~^{\prime\prime })^2}~,
~~~~
E_2^\prime =\sqrt{m^2+(\vec Q +\vec q~^{\prime\prime })^2}~,
~~~~
E_3^\prime=\sqrt{m^2+\vec q~^{\prime\prime 2}}~~.
\end{eqnarray}

Applying Eqs.(\ref {daxis})-(\ref {angles}) to describe the transformation
of the deuteron wave function due to the Wigner rotation, we 
can find the directions of the rotation axes for the initial
and final deuterons, respectively:
\begin{eqnarray}
\vec n=\frac{\vec q~^\prime\times\vec Q}{|\vec q~^\prime\times\vec Q|}~,
~~~~~~~~
\vec n^\prime=\frac{\vec q~^{\prime\prime}\times\vec Q}
{|\vec q~^{\prime\prime}\times\vec Q|}~~.
\end{eqnarray}
Note, these  axes  do not coincide here
as it was in the one nucleon exchange and  single scattering cases.

The rotation angles are expressed through their tangents, which are defined 
by Eq.(\ref{tan}) for the initial deuteron, and by the following relations $-$
for the final deuteron:
\begin{eqnarray}
tg\frac{\omega_2^\prime}{2}&=&-\frac{|\vec q~^{\prime\prime }\times \vec Q |}
{m(E_2^\prime +E_3^\prime+2E^{\prime *})+2E^{\prime *}(E_2^\prime +E^{\prime *})}
\\
tg\frac{\omega_3^\prime}{2}&=&\frac{|\vec q~^{\prime\prime }\times \vec Q |}
{m(E_2^\prime +E_3^\prime+2E^{\prime *})+2E^{\prime *}(E_3^\prime +E^{\prime *})}~~.
\nonumber
\end{eqnarray}

In Eq.(\ref {ds}) we have two $NN$ $t$-matrices which correspond to two
rescattering vertices. There are the off-energy shell matrices defined
in the previous subsection. We get $\sqrt{s_{13}}$ and $\sqrt {s_{12}^\prime}$ as the arguments
of the first and second $t$-matrices, correspondingly.
 The c.m. relative momenta of the colliding
and scattering nucleons in the first vertex are related with the
nucleon momenta in the deuteron Breit frame by the following expressions:
\begin{eqnarray}
\vec k&=&\vec Q -
\frac{(\vec p +\vec Q)(E_3+E_{13}^*)+\vec q~^\prime (E_p+E_{13}^*)}{E_p+E_3+2E_{13}^*}
\\
\vec k^\prime &=&\vec q~^{\prime\prime} -
(\vec p +\vec Q-\vec q~^\prime )\frac{E_3^\prime +E_{13}^{\prime *}}
{E_1^\prime +E_3^\prime +2E_{13}^{\prime *}}~~.
\nonumber
\end{eqnarray}
The arguments of the second NN $t$-matrix are
\begin{eqnarray}
\vec \kappa &=&\vec q~^{\prime\prime} -
(\vec p +\vec Q-\vec q~^\prime )\frac{E_2 +E_{1}^*}
{E_1^\prime +E_2 +2E_{12}^*}
\\
\vec \kappa~^\prime &=&-\vec Q -
\frac{(\vec p +\vec Q)(E_2^\prime+E_{12}^{\prime *})+
\vec q~^{\prime\prime} (E_p+E_{12}^{\prime *})}{E_p+E_2^\prime +2E_{12}^{\prime *}}~~,
\nonumber
\end{eqnarray}
where the two-nucleons center-of-mass energies can be expressed by means of the
corresponding Mandelstam variables $s$: 
\begin{eqnarray}
E_{12}^*&=&\sqrt{s_{12}}/2=\frac{1}{2}\sqrt{(E_1+E_2)^2-(\vec p+\vec Q-\vec q~^{\prime\prime })^2}
\nonumber\\
E_{12}^{\prime *}&=&\sqrt{s_{12}^\prime }/2=
\frac{1}{2}\sqrt{(E_p+E_2^\prime )^2-(\vec p+\vec Q-\vec q~^{\prime\prime })^2}
\\
E_{13}^*&=&\sqrt{s_{13}}/2=\frac{1}{2}\sqrt{(E_p+E_3)^2-(\vec p+\vec Q-\vec q~^\prime )^2}
\nonumber\\
E_{13}^{\prime *}&=&\sqrt{s_{13}^\prime }/2=
\frac{1}{2}\sqrt{(E_1+E_3^\prime )^2-(\vec p+\vec Q-\vec q~^\prime )^2}~~.
\nonumber
\end{eqnarray}

The normalization factors $N$ and $N^\prime$ are connected with the
transformations of the nucleon-nucleon $t$-matrices  as follows:
\begin{eqnarray}
N=\sqrt{\frac {E_{13}^*(E_p+E_3)}{2E_p E_3}}
\sqrt{\frac {E_{13}^{\prime *}(E_1+E_3^\prime)}{2 E_1 E_3^\prime }}
\frac{\sqrt{s_{13}}}{E_p+E_3}
cos \frac{\chi}{2} cos \frac{\chi_3}{2} cos \frac{\chi_1}{2}
cos \frac{\chi_3^\prime}{2}
\\
N^\prime =\sqrt{\frac {E_{12}^*(E_1+E_2)}{2E_1 E_2}}
\sqrt{\frac {E_{12}^{\prime *}(E_p+E_2^\prime)}{2 E_p E_2^\prime }}
\frac{\sqrt{s^\prime_{12}}}{E_p+E_2^\prime}
cos \frac{\varphi_1}{2} cos \frac{\varphi_2}{2}
cos \frac{\varphi}{2}
cos \frac{\varphi_2^\prime}{2}~~.
\end{eqnarray}
Each of these factors includes the normalization coefficient
${\cal N}$ related with the Jacobians of the transformations
from the c.m. to the reference frame, Eq.(\ref {Jac}), 
the kinematical factor $F$  defined in Eq.(\ref {F}),
and cosines of the Wigner rotation angles which  follow from
the Wigner rotation operators, Eq.(\ref {w12}). The expressions for 
the tangents of the Wigner angles are given in Appendix.

The three nucleon free propagator in Eq.(\ref {ds}) can be decomposed on the two
terms using the well known formula: 
\begin{eqnarray}
\frac{1}{E_d+E_p-E_1^\prime -E_2^\prime -E_3^\prime +i\varepsilon}&=&
{\cal P}\frac{1}{E_d+E_p-E_1^\prime -E_2^\prime -E_3^\prime }-
\nonumber\\
\\
&&
i\pi\delta(E_d+E_p-E_1^\prime -E_2^\prime -E_3^\prime )~~.
\nonumber
\end{eqnarray}
The principal value part is often neglected to simplify further calculations.
However this method is efficient only to evaluate the double  scattering
contribution. In order to get a more precise result, it is very important 
to take into account the both terms of the
three nucleon propagator .
In our calculations we perform all integrations without any simplifications by
using standard CERNLIB programs. The difference between the results obtained
with and without the principal value part of the free propagator, is
discussed below. 

\section{Polarization observables}

In the previous section the amplitude of the deuteron- proton
elastic scattering has been considered in detail.
We have shown how  three contributions contained in this 
amplitude can be expressed in terms of the deuteron wave
function and nucleon-nucleon $t$-matrices defined in the deuteron
Breit frame.

On the other hand, as  in ref. \cite{alberi},
the $dp$ elastic scattering amplitude ${\cal J}_{dp\to dp}$ 
can be decomposed in spin-1 operators acting in the deuteron
spin space and spin-1/2 operators $\sigma_i$ acting in the proton
spin space: 
\begin{eqnarray}
{\cal J}_{dp\to dp}&=&<m^\prime {\cal M}_d^\prime |f_1+f_2(\vec S\vec y) +
f_3 Q_{xx}+f_4 Q_{yy}+f_5 (\vec\sigma\vec x)(\vec S\vec x)+
\nonumber\\
&&
\hspace{1.7cm}
f_6(\vec\sigma\vec x) Q_{xy}+
f_7(\vec\sigma\vec y)+
f_8(\vec\sigma\vec y)(\vec S\vec y)+f_9 (\vec\sigma\vec y)Q_{xx}+
\\
&&\hspace{1.7cm}
f_{10}(\vec\sigma\vec y) Q_{yy}+
f_{11}(\vec\sigma\vec z)(\vec S\vec z)+
f_{12}(\vec\sigma\vec z)Q_{yz}|m {\cal M}_d>~~.
\nonumber
\end{eqnarray}
 Due to
 the time-reversal and parity invariance, we have only 12 linearly 
 independent amplitudes $f_i$. These amplitudes can be calculated by
 using the technique presented above.
The operators $\sigma_i$ are the Pauli
 matrices, while  $S_i$   and $Q_{ij}$ are the
 spin-1 vector and quadrupole operators, respectively:
\begin{eqnarray}
Q_{ij}=\frac{1}{2}(S_i S_j+S_j S_i)-\frac{2}{3}\delta_{ij} \hat I~,
~~~~~~~~
Q_{xx}+Q_{yy}+Q_{zz}=0~~.
\end{eqnarray} 

This paper  considers only two polarization observables:
the vector analyzing power of the deuteron $A_y$ and tensor
analyzing power $A_{yy}$ which are defined
by the standard manner:
\begin{eqnarray}
A_y=\frac{Tr({\cal J} S_y {\cal J}^\dagger)}{Tr({\cal J J}^\dagger)}~,
~~~~~~~~
A_{yy}=\frac{Tr({\cal J} Q_{yy} {\cal J}^\dagger)}{Tr({\cal J J}^\dagger)}~~.
\end{eqnarray}
 
The squared $dp$-amplitude summarized over all the spin projections can be presented
through amplitudes $f_i$ as 
\begin{eqnarray}
Tr({\cal J J}^\dagger)&=&6 (f_1^2+f_7^2)+4 (f_2^2+f_5^2+f_8^2+f_{11}^2)+
\frac{4}{3} (f_3^2+f_4^2+f_9^2+f_{10}^2)+
\\
&&
(f_6^2+f_{12}^2)-
\frac{4}{3} Re(f_3 f_4^*+f_9f_{10}^*)~~.
\nonumber
\end{eqnarray}
Using the standard technique one can also obtain   
traces to define
 the vector
\begin{eqnarray}
Tr({\cal J} S_y {\cal J}^\dagger)&=&Re \Bigg [ 
8 (f_1 f_2^*+f_7 f_8^*)-\frac{4}{3}(f_2 f_3^*+f_8 f_9^*)+
\\
&&\hspace{0.8cm}
\frac{8}{3} (f_2 f_4^*+f_8 f_{10}^*)+2(f_5 f_6^*+f_{11} f_{12}^*)\Bigg ]
\nonumber
\end{eqnarray}
and tensor analyzing powers
\begin{eqnarray}
Tr({\cal J} Q_{yy} {\cal J}^\dagger)&=&
\frac{4}{3}(f_2^2+f_8^2)+\frac{2}{9}(f_3^2+f_9^2)-
\frac{4}{9}(f_4^2+f_{10}^2)
-\frac{2}{3}(f_5^2+f_{11}^2)-
\nonumber\\
&&\frac{1}{6}(f_6^2+f_{12}^2)+
\frac{8}{3} Re (f_1 f_4^*+f_7 f_{10}^*)-
\frac{4}{3} Re (f_1 f_3^*+f_7 f_9^*)+
\\
&&\frac{4}{9} Re (f_3 f_4^*+ f_9 f_{10}^*)~~.
\nonumber
\end{eqnarray}

In general, we have to apply the spin transformation technique
to relate the laboratory observables to the Breit frame ones.
However, we have considered only those observables which are defined
by the polarization along the normal to the scattering plane.
In this case the definitions of the polarization observables
are identical both $-$ for the laboratory and Breit frames. It is not
the case for other observables, for example, $A_{xx}$, $A_{xz}$, etc.

\section{Results and discussions}

The results of our calculations are presented in Figs.(2-5).
We have performed our investigations at two deuteron kinetic 
 energies: 395 MeV and
1200 MeV. This choice was motivated by
two reasons. First, there are  sets of experimental data at
these energies \cite {395}-\cite {1200a}.  Second, these energies have
 demonstrated
the range, where the  method presented here can be applied.

The dashed curves in Figs. (2-5) correspond to the calculations which
include only the one nucleon exchange and single scattering (ONE+SS)
contributions. The solid line is the result which also takes the double
scattering (DS) into account. All the calculations have been carried out
 with the
CD Bonn deuteron wave function \cite {cd}.
We have  obtained a good agreement of the theoretical
curves with the experimental data at the deuteron kinetic energy of 395 MeV.
Note, the contribution of the double scattering is not very
big at this energy, especially for $A_{yy}$.

The other situation has been revealed for the higher energy.
 An agreement  between the
obtained curves and experimental data at the deuteron energy of 1200 MeV
is not so obvious. The vector analyzing power is very well
described up to the angle of  $90^0$ c.m., while the behaviour
of the tensor analyzing power is reproduced only qualitatively in a wide
 angular range. Here the single scattering contribution dominates
only up to $\theta ^*=30^0$, whereas the both polarization observables 
are very sensitive at this energy to the double scattering at larger
angles. Especially
the behaviour of the $A_{yy}$ changes very strongly due to 
the DS term having been taken into account. On the contrary,
 at the deuteron energy of 395 MeV
the double scattering contribution is 
practically insignificant for the tensor analyzing power but
 for the vector analyzing power it is up to $50\%$.

The results of the calculations with and without the principal value part
of the three-nucleon free propagator are given in Figs.(6-9).
The curves corresponding to the predictions for the
deuteron energy of 395 MeV, are practically undistinguished.
It is a rather clear result, since the double scattering contribution
 is not so large at this energy. 

At 1200 MeV the contribution of the principal value part of the free
propagator is more significant. Note, that the full calculation describes
the vector analyzing power better than the simplified method, especially
at the $40^0-80^0$ c.m. angles, where the double scattering is very
important. But it is not the case for the tensor analyzing power.
Here the full calculation gives the smaller absolute value
as compared with the experimental data in the range, where $A_{yy}$
has a maximum or minimum , while
the result obtained by the reduced method is in a good agreement with the data.
 However, it is necessary to keep in mind, that this shortage is observed even 
in the range,
where the single scattering  dominates ($\theta^*\le 40^0$).
From the above it is possible to conclude, that the reason of this deficiency is not
related with  the three nucleon free propagator in
the double scattering term.

It should be noted that these observables are also sensitive  to the
nucleon-nucleon amplitudes. Unfortunately, to describe $NN$
interactions we have used the model offered twenty years ago, when the
nucleon-nucleon phase shift analysis was incomplete. Besides,
up to now we have no reliable data of the $np$ amplitudes at the nucleon
kinetic energy higher than 1.3 GeV \cite {said} and that is very important 
to study
reactions with light nuclei at intermediate energies. Maybe, that
the qualitative disagreement between the theoretical predictions
and experimental data at the deuteron energy of 1200 MeV is caused by
incomplete
 $NN$ interaction description.
In this connection the modern parameterization of the nucleon-nucleon
$t$-matrix in a wide energy range is needed.

\section{Conclusion}

In this paper we have presented a method to calculate 
the amplitude of the deuteron-proton elastic scattering at intermediate
energies. Special attention was given to the questions connected
with the relativistic effects. The transformation of the deuteron
wave function in the rest frame to a moving system was performed, that
 allowed us to use the nonrelativistic DWF
  at rather  high energies.
In order to describe nucleon-nucleon interactions in a wide energy range,
we have used
 parameterization of the $NN$ $t$-matrix. The spin transformation
technique has been also applied to relate this $t$-matrix given in the c.m.
to that in the reference frame.

Using the method  presented here we have managed to describe the experimental
data on the vector, $A_y$, and tensor, $A_{yy}$, analyzing powers at
two energies. We got a good agreement between our predictions and
experimental data at the deuteron kinetic energy of 395 MeV and acceptable
agreement $-$ at 1200 MeV. We believe that the difference between the obtained
predictions and data is related with the shortcomings of the 
nucleon-nucleon parameterization. It is possible that the deuteron-proton
scattering description can be improved, if the modern $NN$ model is
 used.

In the future it is planned to apply the approach developed in this paper
to describe experimental data at other energies. In particular,
the data obtained at the Nuclotron (Dubna) at 880 MeV of the deuteron 
energy \cite {lad}
can be described in this formalism. 

\vspace{1cm}
{\it Acknowledgements:}
The author is grateful to Dr. V.P. Ladygin for fruitful discussions.
This work has been supported by the Russian Foundation for Basic Research
under grant  $N^{\underline 0}$  07-02-00102a.

\vspace{1cm}
{\bf\large Appendix}

Applying Eq.(\ref {w12}) we can obtain the expressions for the Wigner
angles. For the first nucleon-nucleon scattering we have 
\begin{eqnarray}
tg\frac{\chi}{2}&=&\frac{|\vec p\times \vec p_3|}{m(E_p+E_3+2E_{13}^*)+2E_{13}^*(E_p+E_{13}^*)}
\nonumber\\
tg\frac{\chi_3}{2}&=&\frac{|\vec p\times \vec p_3|}{m(E_p+E_3+2E_{13}^*)+2E_{13}^*(E_3+E_{13}^*)}
\nonumber\\
tg\frac{\chi_1}{2}&=&\frac{|\vec p_1\times \vec q^{\prime\prime}|}
{m(E_1+E_3^\prime+2E_{13}^{\prime *})+2E_{13}^{\prime *}(E_1+E_{13}^{\prime *})}
\nonumber\\
tg\frac{\chi_3^\prime }{2}&=&\frac{|\vec p_1\times \vec q^{\prime\prime}|}
{m(E_1+E_3^\prime+2E_{13}^{\prime *})+2E_{13}^{\prime *}(E_3^\prime+E_{13}^{\prime *})}~~,
\nonumber
\end{eqnarray}
where $\chi $, $\chi_3$ are the angles corresponding to the beam proton
and struck nucleon from the deuteron, respectively, and $\chi_1$, $\chi_3^\prime$ are the 
angles corresponding to the scattered nucleons (Fig. 1c).

The analogous expressions are obtained for the second scattering vertex:
\begin{eqnarray}
tg\frac{\varphi_1}{2}&=&\frac{|\vec p_1\times \vec q^\prime|}
{m(E_1+E_2+2E_{12}^*)+2E_{12}^*(E_1+E_{12}^*)}
\nonumber\\
tg\frac{\varphi_2}{2}&=&\frac{|\vec p_1\times \vec q^\prime|}
{m(E_1+E_2+2E_{12}^*)+2E_{12}^*(E_2+E_{12}^*)}
\nonumber\\
tg\frac{\varphi}{2}&=&\frac{|\vec p^\prime\times \vec p_2^\prime|}
{m(E_p+E_2^\prime +2E_{12}^{\prime *})+2E_{12}^{\prime *}(E_p+E_{12}^{\prime *})}
\nonumber\\
tg\frac{\varphi_2^\prime }{2}&=&\frac{|\vec p^\prime\times \vec p_2^\prime|}
{m(E_p+E_2^\prime +2E_{12}^{\prime *})+2E_{12}^{\prime *}(E_2^\prime+E_{12}^{\prime *})}~~.
\nonumber
\end{eqnarray}
Here $\varphi_1$, $\varphi_2$ correspond to the colliding nucleons, while
$\varphi$, $\varphi_2^\prime$ correspond to the scattered proton and 
nucleon from the deuteron, respectively.

\newpage

{\large\bf Figure captions }

\vspace{2cm}

{\bf Fig.1} The diagrams included into consideration.
(a) The one nucleon exchange diagram. (b) The single
scattering diagram. (c)  The double scattering diagram.

\vspace{1cm}

{\bf Fig.2} The deuteron vector analyzing power at 395 MeV as a function
of the c.m. angle, the data are taken from ref.  \cite {395}.
 The dashed line corresponds
to the ONE+SS, the solid line shows the ONE+SS+DS calculation.  

\vspace{1cm}

{\bf Fig.3} The deuteron tensor analyzing power at 395 MeV as a function
of the c.m. angle, the data are taken from ref. \cite {395}.
 The dashed line corresponds
to the ONE+SS, the solid line shows the ONE+SS+DS calculation.

\vspace{1cm}

{\bf Fig.4} The deuteron vector analyzing power at 1200 MeV as a function
of the c.m. angle, the data are taken: ($\bullet$) from ref. \cite {1200},
($\triangle$) from ref. \cite {1200a}. The dashed line corresponds
to the ONE+SS, the solid line shows the ONE+SS+DS calculation.  

\vspace{1cm}

{\bf Fig.5} The deuteron tensor analyzing power at 1200 MeV as a function
of the c.m. angle, the data are taken: ($\bullet$) from ref. \cite {1200}, 
({$\triangle$) from ref. \cite {1200a}. The dashed line corresponds
to the ONE+SS,  the solid line shows the ONE+SS+DS calculation.

\vspace{1cm}

{\bf Fig.6} The deuteron vector analyzing power at 395 MeV as a function
of the c.m. angle, the data are taken from ref. \cite {395}. 
The solid and dashed lines
 correspond
to the calculations with and without the principal value part of the propagator.  

\vspace{1cm}

{\bf Fig.7} The deuteron tensor analyzing power at 395 MeV as a function
of the c.m. angle, the data are taken from ref. \cite {395}.  
The solid and dashed lines
 correspond
to the calculations with and without the principal value part of the propagator.

\vspace{1cm}

{\bf Fig.8} The deuteron vector analyzing power at 1200 MeV as a function
of the c.m. angle, the data are taken: ($\bullet$) from ref. \cite {1200},
($\triangle$) from ref. \cite {1200a}.  The solid and dashed lines
 correspond
to the calculations with and without the principal value part of the propagator.    

\vspace{1cm}

{\bf Fig.9} The deuteron tensor analyzing power at 1200 MeV as a function
of the c.m. angle, the data are taken: ($\bullet$) from ref. \cite {1200},
($\triangle$) from ref. \cite {1200a}.  The solid and dashed lines
 correspond
to the calculations with and without the principal value part of the propagator.    

\newpage

\begin{figure}[t]

\begin{minipage}{6cm}
 \epsfysize=150mm
 \epsfbox{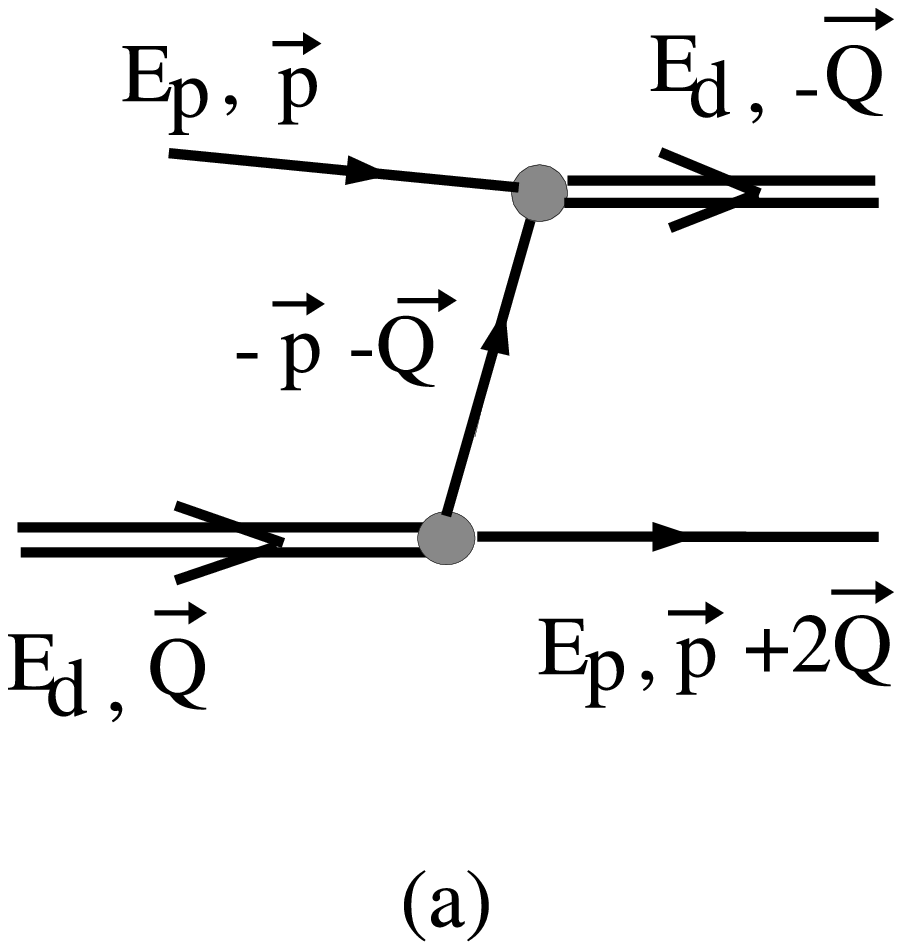}
\end{minipage}
\end{figure}


\begin{figure}[t]
\vspace{-16cm}
\hfill{
\begin{minipage}{6cm}
 \epsfysize=140mm
 \epsfbox{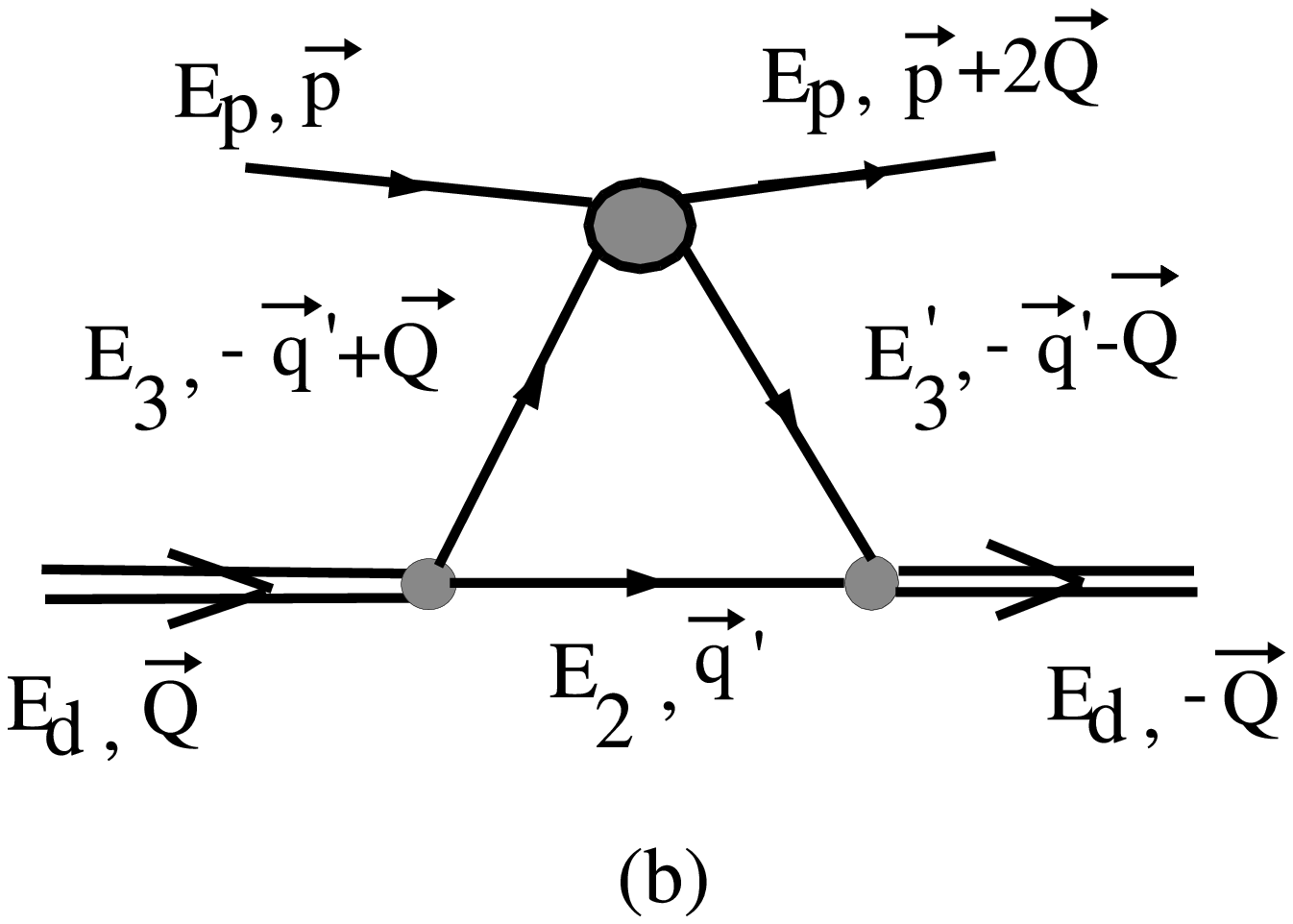}
\end{minipage}
}
\end{figure}

\begin{figure}[t]
\vspace{-8cm}
 \epsfysize=140mm
 \centerline{
 \epsfbox{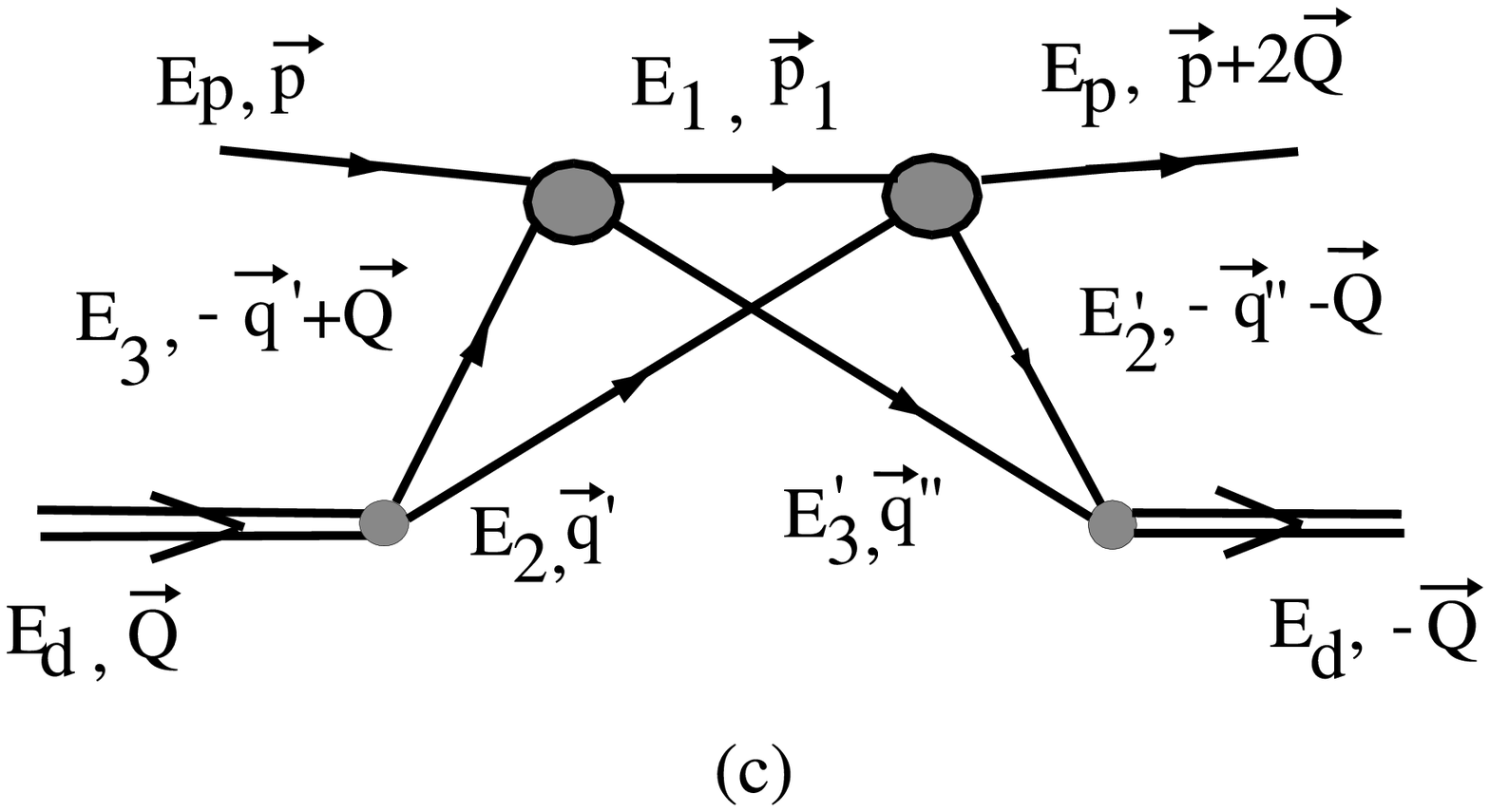}
 }
 \caption{Fig.1 }
\end{figure}

\newpage

\begin{figure}[h]

 \centerline{
 \epsfbox{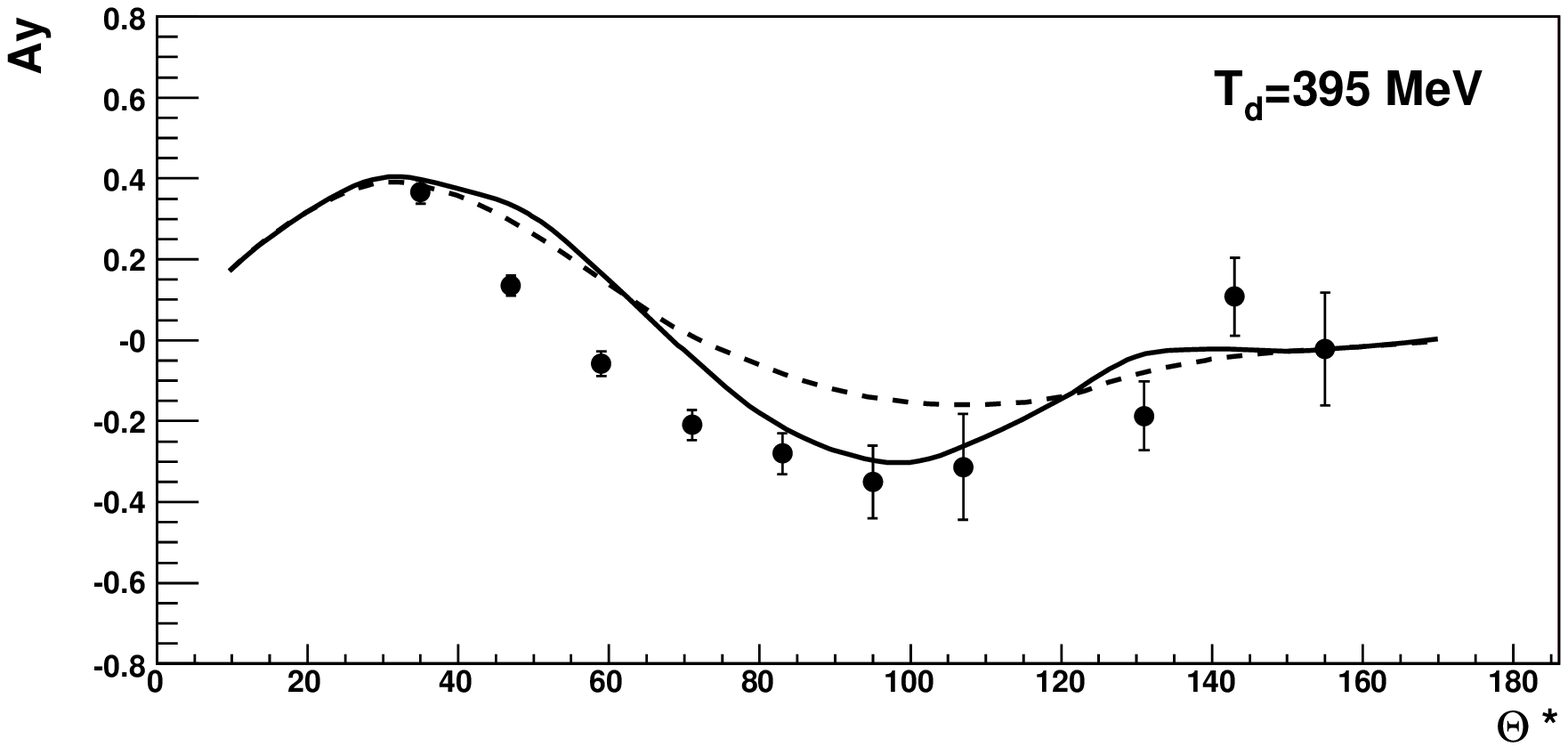}
 }
 \caption{Fig.2 }
\end{figure}

\newpage

\begin{figure}[h]

 \centerline{
 \epsfbox{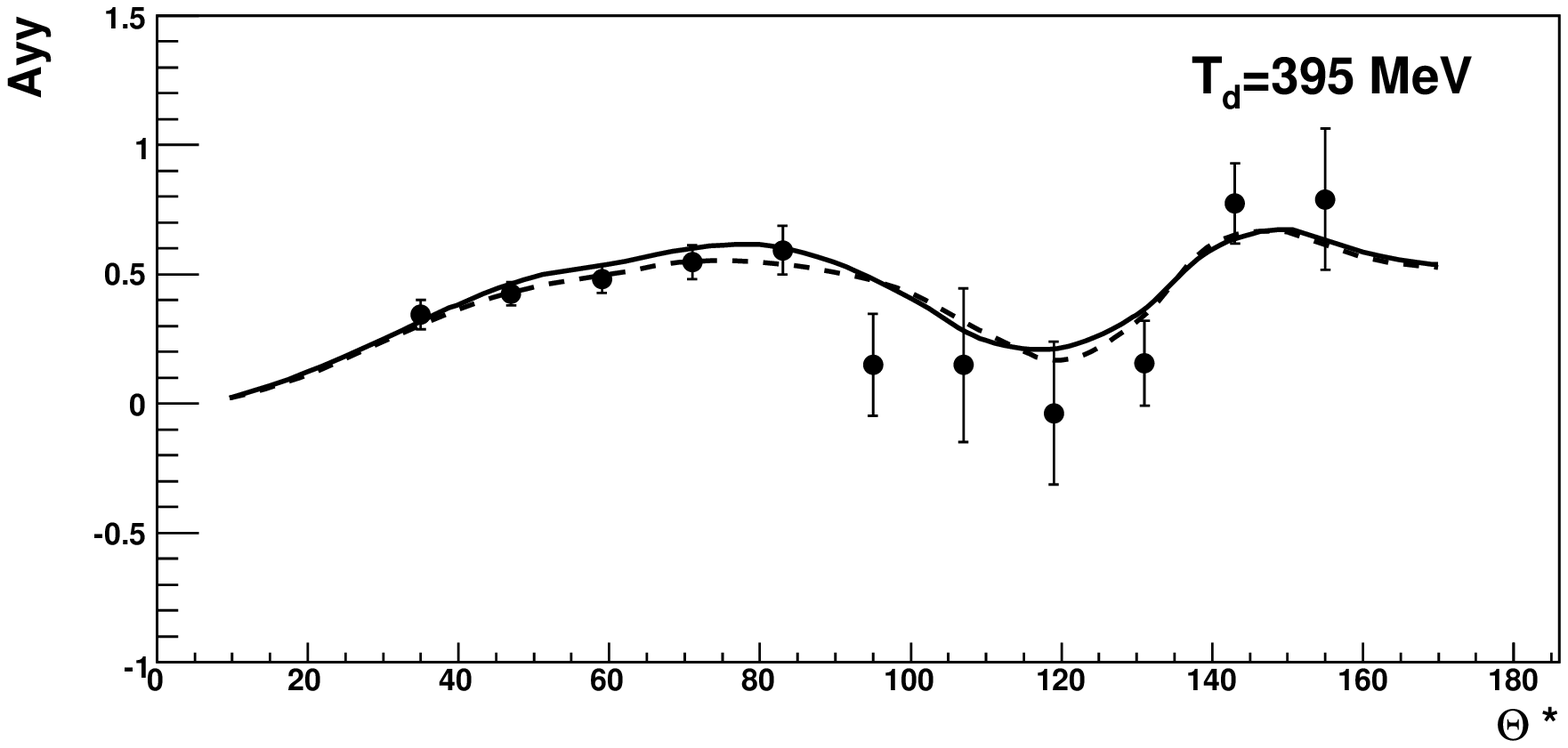}
 }
 \caption{Fig.3 }
\end{figure}

\newpage

\begin{figure}[h]

 \centerline{
 \epsfbox{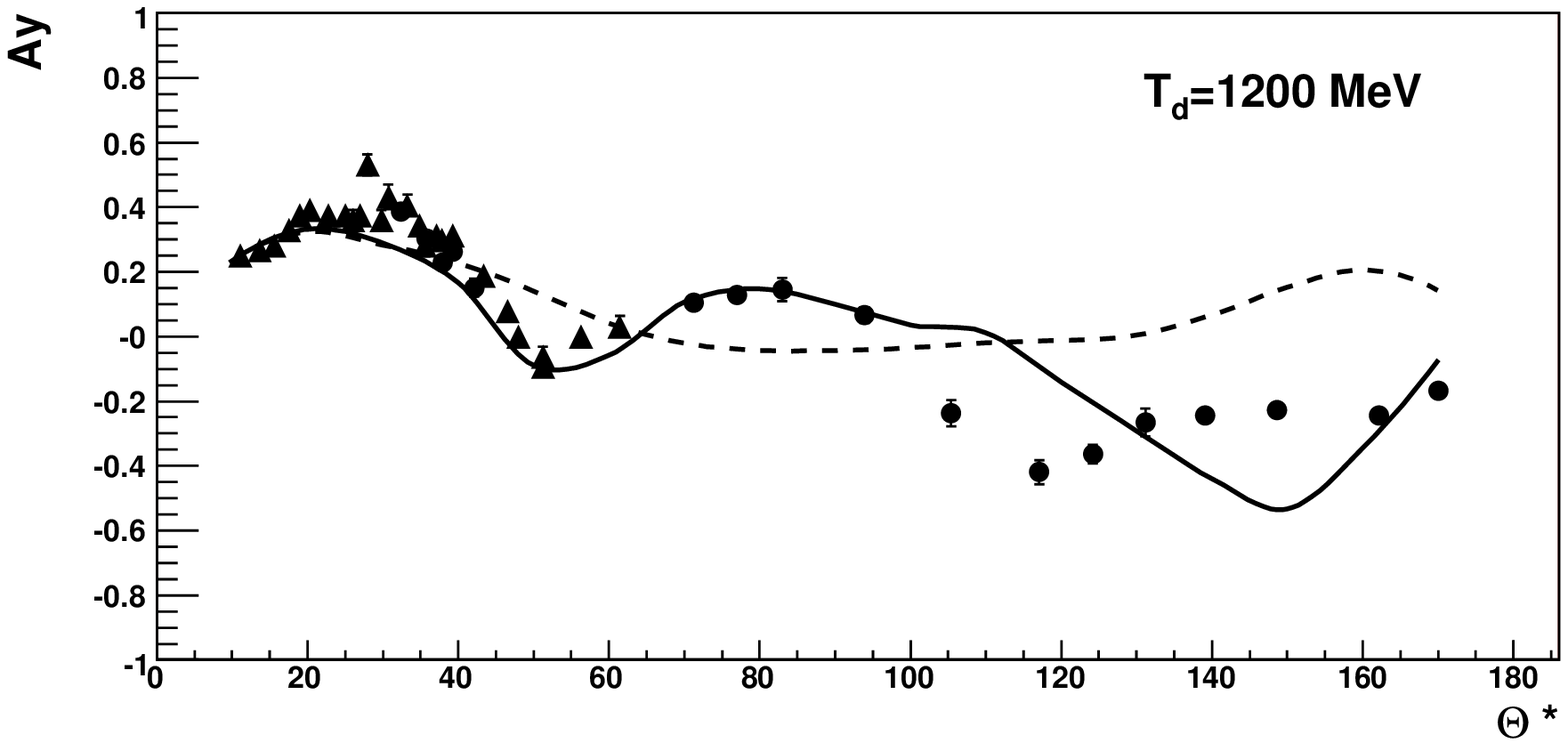}
 }
 \caption{Fig.4 }
\end{figure}

\newpage

\begin{figure}[h]

 \centerline{
 \epsfbox{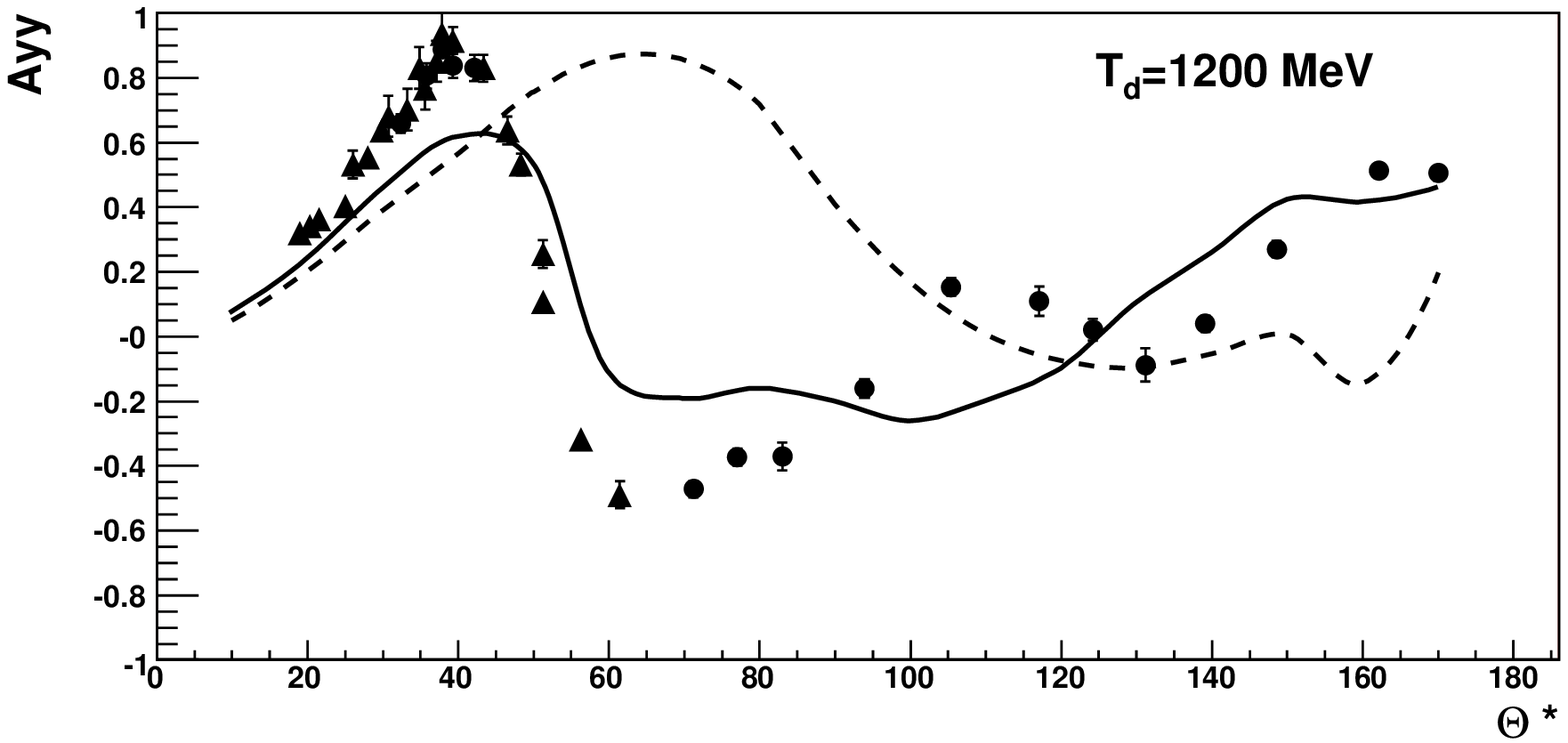}
 }
 \caption{Fig.5}
\end{figure}

\newpage

\begin{figure}[h]

 \centerline{
 \epsfbox{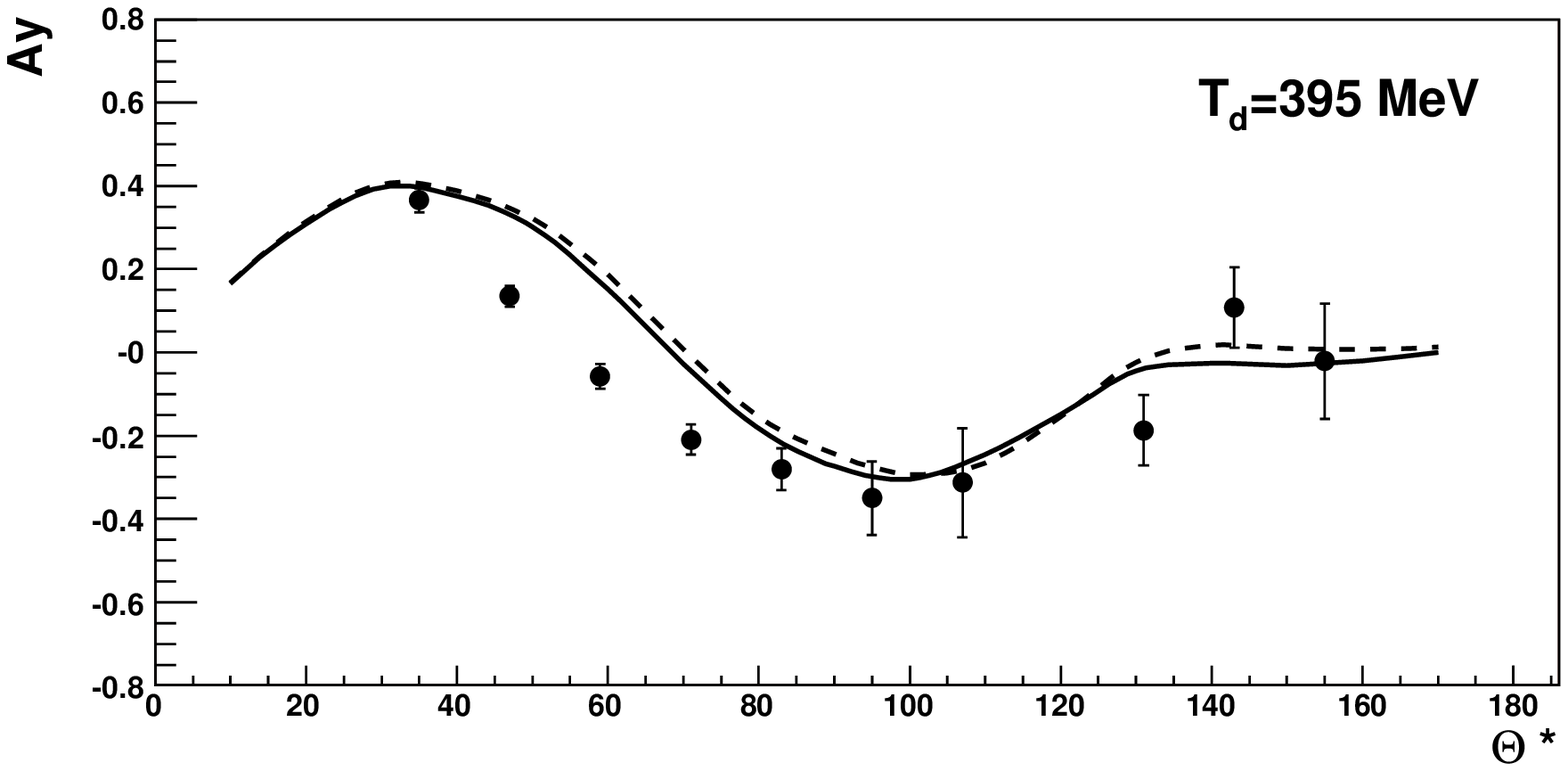}
 }
 \caption{Fig.6}
\end{figure}

\newpage

\begin{figure}[h]

 \centerline{
 \epsfbox{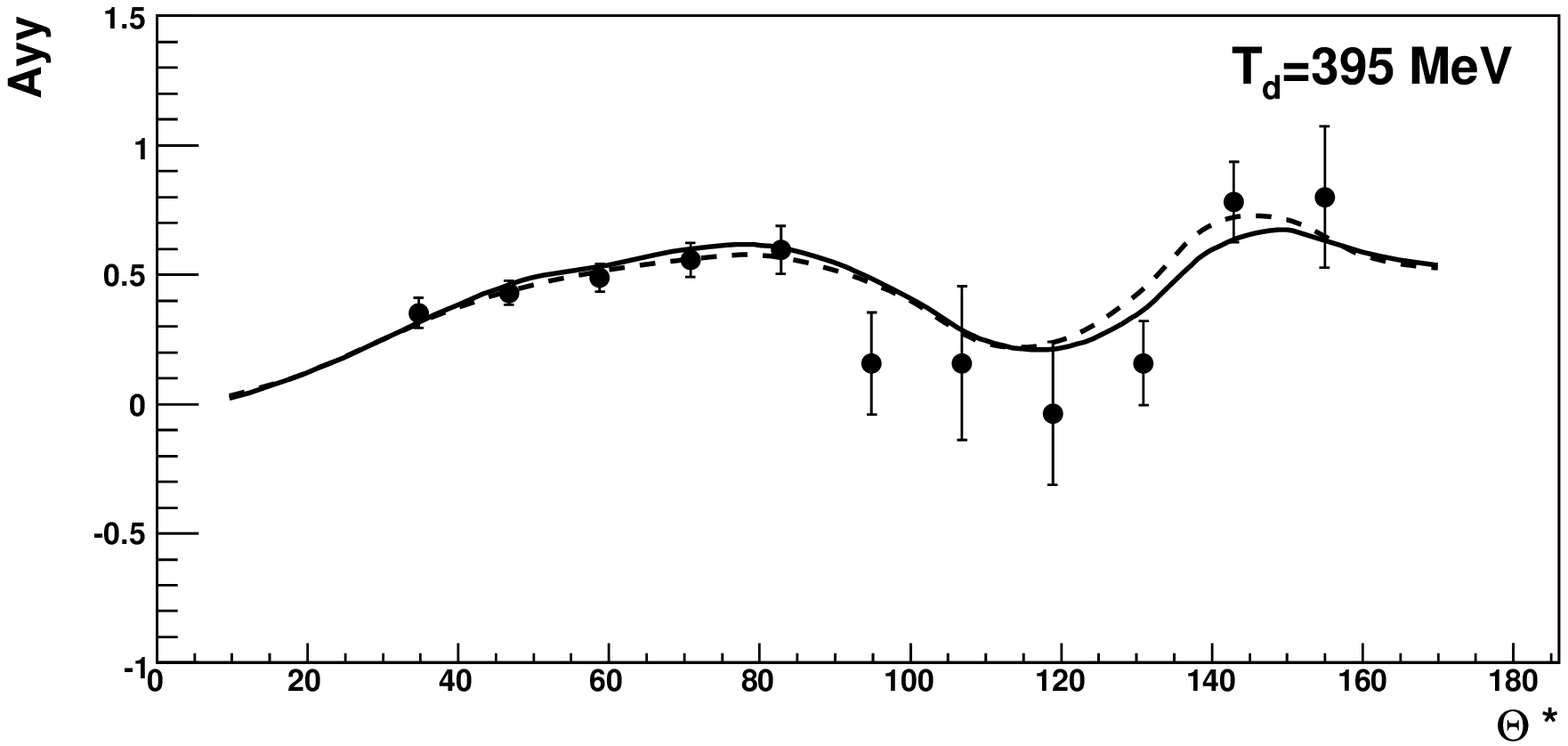}
 }
 \caption{Fig.7}
\end{figure}

\newpage

\begin{figure}[h]

 \centerline{
 \epsfbox{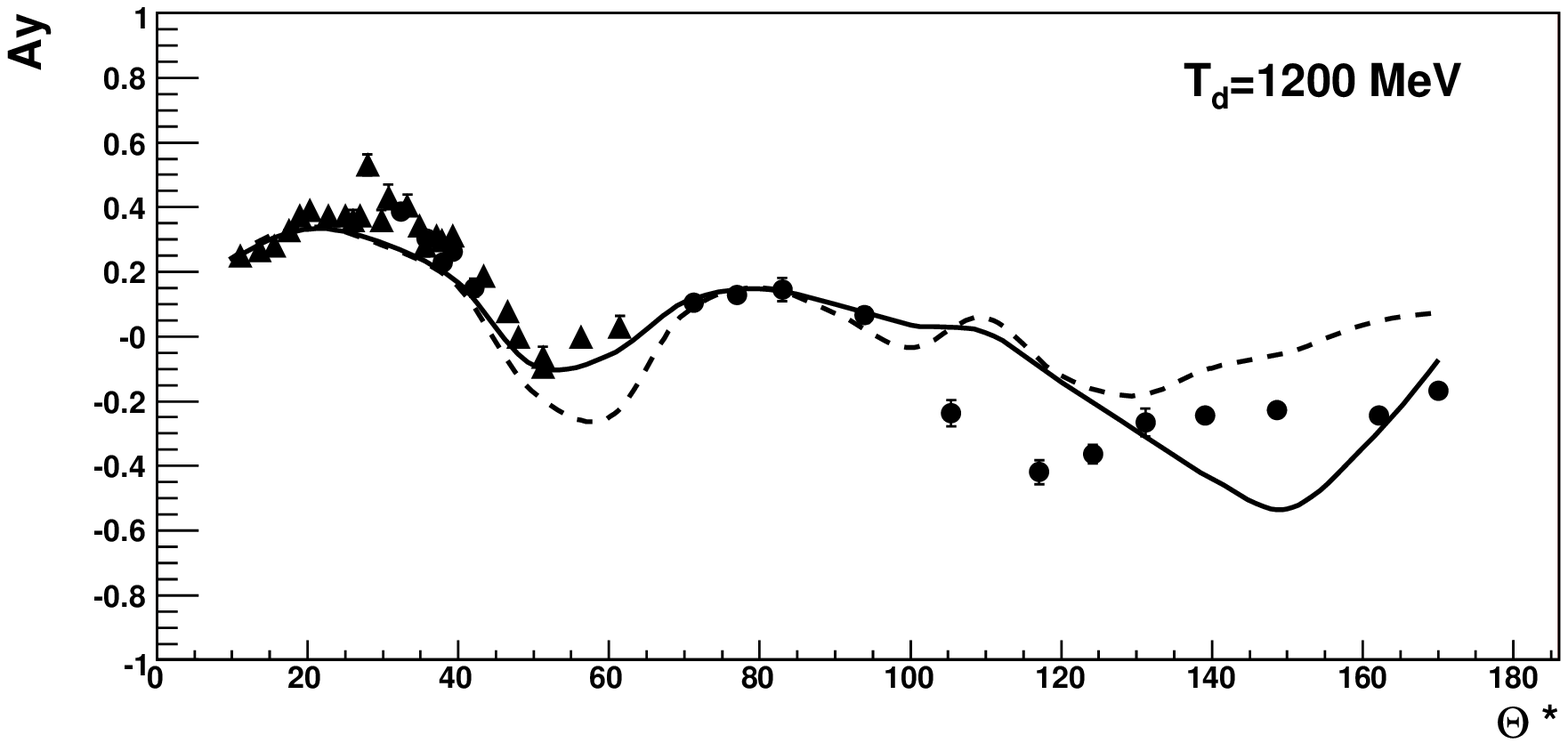}
 }
 \caption{Fig.8}
\end{figure}

\newpage

\begin{figure}[h]

 \centerline{
 \epsfbox{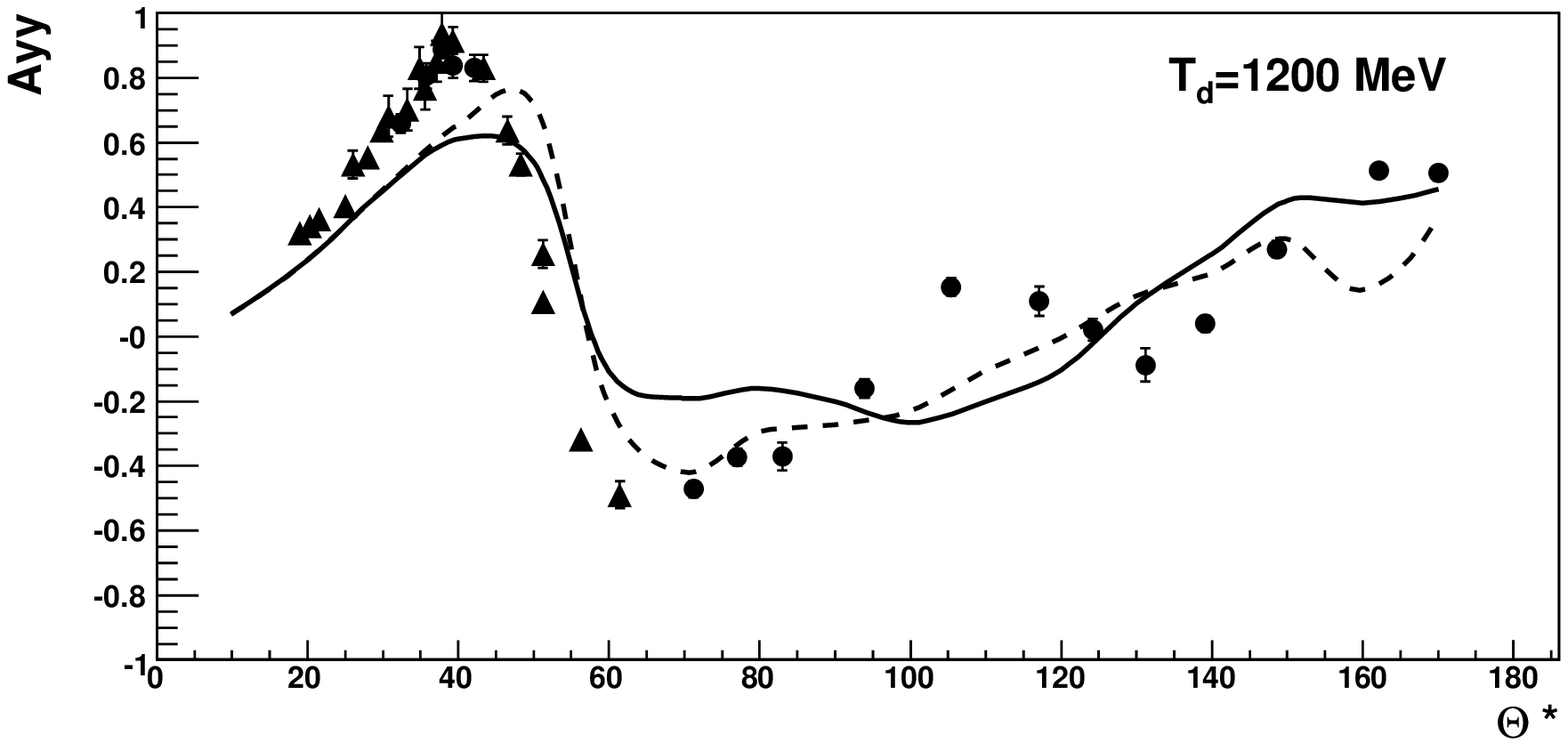}
 }
 \caption{Fig.9}
\end{figure}

\end{document}